\documentclass[natbib]{svmult}
\usepackage{url}

\begin{document}

\sloppy

\newcommand{\ket}[1]{\mbox{$\vert#1\rangle$}}
\newcommand{\bra}[1]{\mbox{$\langle#1\vert$}}
\newcommand{\ketbra}[2]{\ket{#1}\!\bra{#2}}
\newcommand{\mat}[1]{\ketbra{#1}{#1}}
\newcommand{\norm}[1]{\vert #1 \vert}
\newcommand{\Tr}{\text{Tr}}
\newcommand{\psiminus}{\ket{\Psi^-}}
\newcommand{\psiminusl}{\oosrtwo\ket{01}-\oosrtwo\ket{10}}
\newcommand{\squash}[1]{\raisebox{0.04ex}[0pt][0pt]{\small$\textstyle #1$}}
\newcommand{\sqfrac}[2]{\squash{\frac{#1}{#2}}}
\newcommand{\oosrtwo}{\sqfrac{1}{\sqrt{2}}}
\newcommand{\oosrthree}{\frac{1}{\sqrt{3}}}
\newcommand{\oosrtwod}{\frac{1}{\sqrt{2}}}
\newcommand{\oosrthreed}{\sqfrac{1}{\sqrt{3}}}
\newcommand{\pbfrac}[2]{\mbox{$\mbox{}^{#1}\!/_{#2}$}}
\renewcommand{\H}[0]{\mbox{\textsf{H}}}
\newcommand{\U}[0]{\mbox{\textsf{U}}}
\newcommand{\A}[0]{\mbox{\textsf{A}}}
\newcommand{\B}[0]{\mbox{\textsf{B}}}

\hyphenation{between}
\hyphenation{effect}
\hyphenation{instead}
\hyphenation{being}
\hyphenation{Review}
\hyphenation{Erwin}
\hyphenation{follow}
\hyphenation{follows}
\hyphenation{matrix}
\hyphenation{signal}
\hyphenation{Alice}
\hyphenation{super-po-si-tion}
\hyphenation{super-po-si-tions}

\title*{Can Free Will Emerge from Determinism in Quantum Theory?}

\author{Gilles Brassard and Paul Raymond-Robichaud}

\institute{Gilles Brassard, corresponding author \at Universit\'e de Montr\'eal, D\'epartement~IRO,
C.P.~6128, Succ.\ Centre--Ville, Montr\'eal (QC), H3C~3J7~\textsc{Canada}. \email{brassard@iro.umontreal.ca}
\and Paul Raymond-Robichaud \at Universit\'e de Montr\'eal, D\'epartement~IRO,
C.P.~6128, Succ.\ Centre--Ville, Montr\'eal (QC), H3C~3J7~\textsc{Canada}. \email{raymopau@iro.umontreal.ca}}

\authorrunning{G.\ Brassard and P.\ Raymond-Robichaud}

\maketitle

\vspace{-2.5cm}
\begin{raggedleft}
``What is proved by impossibility proofs is lack of imagination'' --- John Bell\\[1ex]
``Imagination is more important than knowledge'' --- Albert Einstein\\
\end{raggedleft}

\vspace{1.5cm}

\newcommand{\abst}{Quantum Mechanics is generally considered to be \emph{the} ultimate theory capable of explaining the emergence of randomness by virtue of the quantum \emph{measurement} process.
Therefore, Quantum Mechanics can be thought of as God's wonderfully imaginative solution to the problem of providing His creatures with Free Will in an otherwise well-ordered
Universe. Indeed, how could we dream of free will in the purely deterministic Universe envisioned by Laplace
if everything ever to happen is predetermined by (and in principle calculable from) the actual conditions
or even those existing at the time of the Big~Bang?\newline\indent
In~this essay, we share our view that Quantum \mbox{Mechanics} is in fact deter\-min\-istic, local
and realistic, in complete contradiction with most people's perception of Bell's theorem,
thanks to our new theory of \emph{parallel lives}. Accord\-ingly,
what we perceive as the \mbox{so-called} ``collapse of the wavefunction'' is but an illusion.
Then we ask the fundamental question: Can a purely deterministic Quantum \mbox{Theory} give rise to
the \emph{illusion} of
nondeterminism, randomness, probabilities, and ultimately can free will emerge from such a theory?}

\abstract*{\abst}

\abstract{\abst}

\keywords{Free will, Realism, Locality, Church of the larger Hilbert space, Determinism, Parallel lives.\\[5mm]
\textbf{To appear in} \textit{Is~science compatible with free will? Exploring free will and consciousness
in light of quantum physics and neuroscience}, A.\ Suarez and P.\ Adams (Eds.), Springer, New York, 2012, Chapter 4.}

\section{Introduction}

By the end of the 19$\mbox{}^{\mbox{\scriptsize th}}$ Century, most physicists had evolved
a completely deter\-min\-istic view of the world.
Even though he had many precursors, such as Paul Henri Thiry, Baron d'\citet*{Holbach},
with his very influential \emph{Syst\`eme de la Nature}, it was the great French mathematician
and astronomer Pierre-Simon, marquis de \citet*{Laplace}, who expressed in the clearest terms the philosophy
\mbox{according} to which everything is predetermined by the initial conditions.
In~his \emph{Essai philosophique sur les probabilit\'es}, he wrote:
\begin{quote}\sf
We ought to regard the present state of the
universe as the effect of its anterior state and as the
cause of the one which is to follow. Given for one
instant an intelligence which could comprehend all the
forces by which nature is animated and the respective
situation of the beings who compose~it---an~intelligence
sufficiently vast to submit these data to analysis---it~would
embrace in the same formula the movements of
the greatest bodies of the universe and those of the
lightest atom; for~it, nothing would be uncertain and
the future, as the past, would be present to its eyes.
\end{quote}

If Laplace were right, would there be any possibility for
conscious \mbox{beings} to exercise free will?
Anything we might imagine that we are deciding would in fact have been ``written'' from the
initial conditions existing at the time of the Big~Bang!
It~would seem that free will requires some form of nondeterminism or randomness;
that it cannot take hold unless some events happen without a cause.\,%
\footnote{Nevertheless, we do acknowledge that \emph{compatibilists} hold the
belief that free will and determinism are compatible ideas, and that it is possible to
believe both without being logically inconsistent.
See~\url{http://plato.stanford.edu/entries/compatibilism/}, accessed on 29~February 2012.}
Even though chaos theory makes it impossible to predict the future in a fully deterministic universe
as soon as there is even the tiniest imprecision on the initial conditions, these initial
conditions would exist precisely according to classical physics, and thus the future would be determined,
independently of our possibility of predicting~it.

In~the 20$\mbox{}^{\mbox{\scriptsize th}}$ Century, quantum mechanics ushered in one of the greatest
revolutions in the history of science.
In~particular, it is generally considered to be \emph{the} ultimate theory capable of explaining the emergence of randomness
by virtue of a mysterious process known as the ``collapse of the wavefunction'', which seems to be inherent to irreversible
quantum \emph{measurements}.
Therefore, quantum mechanics can be thought of as God's wonderfully imaginative solu\-tion to the problem of providing His creatures with free will in an otherwise well-ordered Universe.
Nevertheless, Einstein so disliked~the~idea of~true randomness in Nature that he
claimed to be ``convinced that \emph{He}~[God] does not throw dice'' in a 1926 letter to Born \citep*{EtoB}.
Most physicists \mbox{today} would say that Einstein was wrong in rejecting the occurrence of truly random events.
But~was~he?

In~this essay, we share our view that quantum mechanics is in fact deter\-min\-istic, local
and realistic, in complete contradiction with most people's perception of Bell's theorem,
thanks to our new theory of \emph{parallel lives}. Accord\-ingly,
what we perceive as the \mbox{so-called} ``collapse of the wavefunction'' is but an illusion.
Then we ask the fundamental question: Can a purely deterministic quantum theory give rise to
the \emph{illusion} of nondeterminism, randomness, probabilities, and ultimately can free will emerge from such a theory?

For the sake of liveliness, the nontechnical style of this essay is purposely that of a spontaneous after-dinner speech.
It~is meant for the enjoyment of a curiosity-driven and scientifically-minded 
readership who does not have prior knowledge in quantum mechanics.
Occasional remarks and more rigorous \mbox{details} for the benefit of the expert
are offered in the footnotes with no apologies for the casual reader.
The next three sections review the standard notions of pure and mixed states, of entanglement, and
of how one part of an entan\-gled state can be described. Readers familiar with these
notions may prefer to proceed directly with Sect.~\ref{church}, which describes
the Church of the Larger Hilbert Space, a central notion to this essay since it restores
determinism into quantum mechanics.
Section~\ref{ontic} attempts to go one step further in restoring also locality at the expense
of realism, but it fails to do~so.
Then, Sect.~\ref{parallel} announces our new theory, which we call \emph{parallel lives},
in which both locality and realism are restored in a physical world in which Bell's inequalities
are nevertheless violated.
Finally, Sect.~\ref{freewill} discusses the implication of all of the above on the existence or not of free will,
be it real or illusory.

\section{Pure and mixed states}\label{states}

According to quantum mechanics, one has to distinguish between pure and mixed states.
A~\emph{pure} state, generally denoted \ket{\Psi} following Paul Dirac, is used to represent a state about which
everything is known. For~instance, \ket0 and \ket1 correspond to the classical notion of bits 0 and~1,
whereas \mbox{$\ket{\Psi}= \oosrtwo\ket0+ \oosrtwo\ket1$}
denotes a \emph{qubit} (for ``quantum bit''), whose state is
an equal \emph{superposition} of \ket0 and~\ket1. This means that \ket{\Psi} represents a state that
corresponds \emph{simul\-ta\-neously} to classical bit values 0 and~1, each with \emph{amplitude}~$\oosrtwo$.
If~this qubit is measured in the so-called \emph{computational basis} (\ket0~vs.~\ket1),
standard quantum \mbox{mechanics} has it that it will \emph{collapse} to either classical state~\ket0
or~\ket1, each with a probability given by the square of the norm of the corresponding amplitude,
here \mbox{$\norm{\oosrtwo}^2=\pbfrac12$} for each alternative. Even though the specific result of the measurement
is not determined by the pure state, and two strictly identical particles in that same state could
yield different results following the same measurement, the probabilities associated with
such measurement outcomes are known \mbox{exactly}.
Furthermore, this particular state would behave in a \mbox{totally} deterministic manner if subjected to a \emph{different}
measurement, known in this case as the Hadamard measurement (or~measurement in the Hadamard basis~``\H{}''),
which asks it to ``choose'' \mbox{between}
\mbox{$\H\ket0 = \oosrtwo\ket0+ \oosrtwo\ket1$} and
\mbox{$\H\ket1 = \oosrtwo\ket0 - \oosrtwo\ket1$}.
In~this case, our state would choose the former since~in fact \mbox{$\ket{\Psi}=\H\ket0$}.
\citet*{peres} \emph{defined} a pure state as one for which there exists a complete measurement
(which he calls a ``maximal test'') under which it behaves deterministically.

In~contrast, \emph{mixed} states are used when
there is intrinsic uncertainty not just about the result of some measurement but
about the result of \emph{all} possible complete measurements, hence about the state itself.
One way to picture a mixed state is to think of a black box inhabited by a Daemon.
When a user pushes a button, the Daemon spits out a state that it chooses at random,\,%
\footnote{This must be a \emph{true} random choice, possibly implemented by a quantum-mechanical process;
flipping a classical coin would not suffice here.} say with equal
probability between \ket0 and \ket1. Such a mixed state would be denoted
\begin{equation}\label{comp-mixed} 
\textstyle \mathcal{E}_1 = \{ ( \ket0,\pbfrac12), (\ket1,\pbfrac12) \} .
\end{equation}
More generally, a mixture of $k$ different pure states is denoted
\begin{equation}
\mathcal{E} =
\{(\ket{\Psi_1},p_1), (\ket{\Psi_2},p_2), \ldots, (\ket{\Psi_k},p_k) \} = 
\{(\ket{\Psi_i},p_i)\}_{i=1}^k \, ,
\end{equation}
which means that the
Daemon chooses some \ket{\Psi_i} with probability $p_i$, \mbox{$1 \le i \le k$},
where the probabilities sum up to~1.
It~is legitimate to wonder if such a state is pure since the Daemon knows which \ket{\Psi_i}
it chose,
or if it is mixed since the user does not know.
In~a sense it is both.
Nevertheless, no measurement chosen by the user will provide a deter\-ministic answer.
For instance, a measurement of $\mathcal{E}_1$ in the computational basis will reveal the
Daemon's random choice, which has equal probability \pbfrac12 of being \ket0 or~\ket1.
On~the other hand, a measurement in the Hadamard basis will produce $\H\ket0$ or
$\H\ket1$ with equal probability \pbfrac12 since such would be the case regardless of whether
the Daemon had spit out \ket0 or~\ket1.
Thus we see that the random\-ness lies with the Daemon in one case and with the user's measurement in the other
case, but the final result is the same.
More interestingly, it can be demonstrated that \emph{any} measurement on $\mathcal{E}_1$
that would ask it to choose between two arbitrary one-qubit orthog\-onal states
would choose either one with equal probability.
(Two~states are \emph{orthogonal} if it is possible in principle to distinguish perfectly between them,
such as \ket0 and \ket1, or $\H\ket0$ and $\H\ket1$.)
By~\citeauthor*{peres}' definition, $\mathcal{E}_1$ is not pure since there does not exist a complete measurement under which
it behaves deterministically.

Mixed states can be described as above by a mixture of pure states, but they can also be described in two other ways.
One of them is known as the \emph{density matrix} (aka density \emph{operator}).
This is a matrix (an~array of numbers) that can be calculated mathematically from the more intuitive
mixture \mbox{$\{(\ket{\Psi_i},p_i)\}_{i=1}^k$} of pure states. The remarkable fact about density matrices
is that different mixtures can give rise to the same matrix, yet this matrix represents \emph{all} that is
measurable about the state, by any measurement whatsoever ``allowed'' by quantum mechanics.
For instance, the density matrix that is computed from Eq.~(\ref{comp-mixed}) is identical to that
arising from the apparently different mixture
\begin{equation}\label{comp-mixed-2} 
\textstyle \mathcal{E}_2 = \{ ( \H\ket0,\pbfrac12), (\H\ket1,\pbfrac12) \} .
\end{equation}
In~other words, if we trust a Daemon to send us an equal mixture of \ket0 and~\ket1 (Eq.~\ref{comp-mixed})
but in fact it provides us with  an equal mixture of $\H\ket0$ and~$\H\ket1$ (Eq.~\ref{comp-mixed-2}),
we shall never be able to notice that it is ``cheating''!\,%
\footnote{A~much more remarkable example of cheating is possible for a Daemon
who would be ``paid'' to produce randomly chosen Bell states.  It~could
produce instead pairs of purely classical uncorrelated random bits. These mixtures being identical
in terms of density matrices, such cheating would be strictly undetectable by the user.
This is profitable for the Daemon because classical bits are so much easier to produce
than Bell states!
}
Given that these two mixtures are impossible to distinguish, it makes sense to consider the
corresponding mixed states as actually \emph{identical}.
Just for completeness, notice
that even mixtures featuring more than 2 pure states can be indistinguishable from those above.
For instance, mixture
\begin{equation}\label{comp-mixed-3} 
\textstyle \mathcal{E}_3 = \{ ( \ket0,\pbfrac13), (\frac12 \ket0 + \frac{\sqrt{3}}{2} \ket1,\pbfrac13),
(\frac12 \ket0 - \frac{\sqrt{3}}{2} \ket1,\pbfrac13) \}
\end{equation}
is indistinguishable from (hence identical~to) mixtures $\mathcal{E}_1$ and~$\mathcal{E}_2$
because it gives rise to the same density matrix.

We postpone until  Sect.~\ref{church} discussion of the third way---by~far the most interesting---in~which
one may think of mixed states. 

\section{Entanglement}\label{entanglement}

The concept of \emph{entanglement} was first published (although not named) by \citet*{epr35},
in a failed attempt to demonstrate the \emph{incompleteness} of the quantum formalism.
However, there is historical \mbox{evidence} that the notion had been anticipated by Schr\"odinger several years previously,
who was quick to understand the importance of entanglement:
``I~would not call that \emph{one} but rather \emph{the} characteristic trait of quantum mechanics,
the one that enforces its entire departure from classical lines of thought'' \citep*{Schr35}.
We~could not agree more with this assessment. Our~quantum world is not classical \emph{because}, as spectacularly demonstrated by \citet*{bell64}, entanglement
can\emph{not} be explained by any classical local realistic theory of the sort that was
so dear to Einstein.
(Or~can~it? We'll come back to this question in Sect.~\ref{parallel}\@.)
Indeed, we \mbox{consider} entanglement to be \emph{the} key to understanding Nature.
We~would even go so far as to say that it's our best window into probing the soul of the Universe.
The~other nonclassical aspects of quantum mechanics, such as the quantization of energy
and its consequence on the photoelectric effect---which earned Einstein his Nobel Prize in~1921---are~no
doubt important, but lag far behind the magic of entanglement on our personal wonder scale.

Entanglement is a phenomenon by which two (or~more) physically separated systems
must sometimes be thought of as a single (nonlocal) entity. The~simplest example of entanglement is known
as the \emph{singlet state},
\begin{equation}\label{EPR}
\textstyle \psiminus = \oosrtwod\ket{01}-\oosrtwod\ket{10} \, ,
\end{equation}
which~consists of two particles. A~measurement of both particles
in the computational basis results in either outcome
\ket{01} or~\ket{10}, each with equal probability since \mbox{$\norm{\pm\oosrtwo}^2=\pbfrac12$}.
Here, outcome \ket{01} means that the first particle is measured as \ket0 and the second as~\ket1,
and similarly for outcome~\ket{10}. In~other words, the two particles yield opposite answers
when they are measured in the computational basis.
So~far, this is not more mysterious than if someone had flipped a penny, sliced it through its edge,
put each half-penny in an envelope, and mailed the envelopes to two distant locations.
When the envelopes are opened (``measured''), there is no surprise in the
fact that each one reveals a seemingly random result (heads or tails) but that the two \mbox{results} are
complementary.
Such an explanation would correspond to purely classical \emph{mixed} state
\begin{equation}\label{fake-EPR} 
\textstyle \mathcal{E}_0 = \{ ( \ket{01},\pbfrac12), (\ket{10},\pbfrac12) \}  ,
\end{equation}
where \ket0 stands for \textsf{heads} and \ket1 stands for \textsf{tails}.

What makes this singlet state so marvellous is that quantum mechanics asserts that
$\psiminus$ is indeed the pure state given in Eq.~(\ref{EPR}) and not the mixed state of
Eq.~(\ref{fake-EPR}), and that those are very different indeed.
In~particular, the result of \emph{any} measurement is \emph{not} predetermined (as~it would be with the
half-penny analogy): it~comes into existence only as a result of the measurement itself.
This is particularly mysterious when the two particles are arbitrarily far apart because it is as if
they were magic coins which, when flipped, always provide opposite, yet \emph{freshly} random, outcomes.
In~fact, the two particles provide
opposite answers to \emph{any} complete measurement,
provided they are subjected to the same one.
It's~like an old couple who disagrees on any question you may ask them\ldots{}
even when they don't have a clue about the answer and hence respond randomly!
This phenomenon can be ``explained''
by elementary linear algebra, according to which state~$\psiminus$, as given in Eq.~(\ref{EPR}),
is \emph{mathematically} equivalent to
\begin{equation}\label{EPR2}
\textstyle \psiminus = \oosrtwod \ket{\psi}\ket{\phi} -\oosrtwod \ket{\phi}\ket{\psi}
\end{equation}
for \emph{any} two one-qubit
orthogonal states \ket{\psi} and~\ket{\phi},
such as $\H\ket0$ and~$\H\ket1$.  It~is important to understand that
this behaviour would \emph{not} occur with the mixed state of Eq.~(\ref{fake-EPR})
because, in that case, asking the two particles to ``choose'' between $\H\ket0$ and $\H\ket1$
would produce two random and \emph{uncorrelated} outcomes.

An~entangled state such as the singlet behaves exactly \emph{as if} the first particle, when asked by a
measurement to choose between orthogonal states \ket{\psi} and~\ket{\phi}, flipped a fair coin to decide
which one to select, and then ``\mbox{instructed}'' the other particle to \emph{instantaneously} assume the
opposite state. This gives the \emph{impression} of instantaneous action at a distance, a concept that so
\mbox{revolted} Einstein
that he derisively called it \emph{spuk\-hafte Fern\-wir\-kun\-gen}
(``Spooky action at a distance'').
But is this really what happens or is it only a na\"{\i}ve ``explanation''?
We~shall come back to this most fundamental issue in Sects.~\ref{ontic}
and~\ref{parallel}. 

It has been experimentally demonstrated that if indeed the particles had to communicate,
then the effect of the first measurement on the second particle would have to take place
at least ten thousand times faster than
at the speed of light \citep*{spooky}. Even more amazingly, \emph{relativistic} experiments have been performed,
following a fascinating theoretical proposal by \citet*{SuSc},
in which the
predictions of quantum mechanics continue to hold even when the two particles move apart quickly enough
that they are both measured before the other in their respective inertial reference frames \citep*{spacelike}.
These remarkable experiments make
it untenable to claim that the first measured particle somehow sends a signal to tell the other how to behave.
This has prompted \citet*{nicolas} to assert that quantum correlation ``emerge from outside space-time''.

We~highly recommend the exceptionally lucid and entertaining popular \mbox{accounts} of some classically-impossible
marvels made possible by entanglement that have been written by \citet*{mermin81,mermin94}
for the \textit{American Journal of Physics}.

\section{Describing one part of an entangled state}

The defining characteristic of a pure entangled state split between two distant locations
is that neither of the local subsystems can be described as a pure state of its own.
This should be clear from \citeauthor*{peres}' definition of a pure state and the fact that each part of an entangled
state is so that its outcome is not predetermined, no matter to which complete measurement it is subjected.
Never\-theless, it makes sense to wonder if there is a way to describe the state of one of the subsystems. 

One natural approach is to see what would happen if we measured the \emph{other} subsystem.
Consider for instance the singlet state $\psiminus$ and let us measure one of the particles in the computational basis.
We~have seen that the outcome is \ket0 (resp.~\ket1) with probability~$\pbfrac12$, in which case the other system is now in
state~\ket1 (resp.~\ket0). Therefore, if one system is measured \emph{and one forgets the outcome of the measurement},
the unmeasured system is in state \ket0 with probability~\pbfrac12 and in state \ket1 also with probability~\pbfrac12.
In~other words, this system is in mixed state~$\mathcal{E}_1$, according to Eq.~(\ref{comp-mixed}).
But~the first system could have been measured in the Hadamard basis instead.
\mbox{Depending} on the result of this measurement, the unmeasured system would then be left either in state
$\H\ket0$ or~$\H\ket1$, each with probability~\pbfrac12. If~we forget again the result of the measurement,
the unmeasured system is therefore in mixed state~$\mathcal{E}_2$, according to Eq.~(\ref{comp-mixed-2}).
Now,~remember that mixtures $\mathcal{E}_1$ and~$\mathcal{E}_2$ are
considered to be identical
since they give rise to the same density matrix. More generally,
it~can be demonstrated from the formalism of quantum mechanics that no matter which complete measurement is performed on one subsystem of an arbitrary entangled state,
the other subsystem always ends up in the same mixed state in terms of a density matrix,
albeit not necessarily according to the same mixture of pure states.

It follows that nothing can be more natural than to describe one subsystem of an entangled state
by the mixed state
in which this subsystem \emph{would} be left \emph{if} the \emph{other} subsystem were measured.
This is well-defined since
the resulting density matrix does not depend on
how the other subsystem is measured.
If~we carry this reasoning to its inescapable conclusion,
it makes sense to describe the state of a subsystem
in this way \emph{even if the other subsystem has not been measured yet}, indeed even if
it is \emph{never} to be measured. When we consider the state of a subsystem of an entangled state
in this way, we say that we \emph{trace out} the other subsystem.

Section~\ref{entanglement} may have left you with the impression that entanglement
\mbox{requires} instantaneous communication, which would be incompatible with Einstein's special theory of relativity.
If~we remember that the density matrix \mbox{describes} all that can be measured about a quantum system,
however, it follows from the above discussion that entanglement can\emph{not} be used to \emph{signal} information
between two points in space since no operation performed on one subsystem of an entangled state
can have a \emph{measurable} effect on the other subsystem.
It~is as if quantum systems were capable of instantaneous communication, but only in tantalizing ways
that could not be harnessed by us, macroscopic humans, to establish such communication between ourselves.
We~shall come back on the consequences of this crucial issue in Sects.~\ref{ontic} and~\ref{parallel}.

\section{Church of the Larger Hilbert Space}\label{church}

We have just seen that the state of any subsystem of a pure (or,~for that matter, mixed as well)
entangled state can be expressed as a mixed state in a unique and natural way.
It~is remarkable that the converse holds.
We~saw in Sect.~\ref{states} that mixed states can be described either as mixtures of pure states
(possibly under the control of a Daemon) or~as density matrices, but we promised a third way and here it~is.
\emph{Any} mixed state can be described as the trace-out of some subsystem from an
appropriate \emph{pure} state. Such a pure state
is called a \emph{purification} of the mixed state under consideration.

There is an easy way (theoretically speaking) to~construct a purification of an arbitrary
mixture  \mbox{$\mathcal{E}=\{(\ket{\Psi_i},p_i)\}_{i=1}^k$}.
For~this, consider some other quantum system that could be in any of $k$ orthogonal states
$\ket{\Phi_1}$, $\ket{\Phi_2}$, \ldots, $\ket{\Phi_k}$ 
and consider pure state
\[ \ket{\Psi} = \sum_{i=1}^k \sqrt{p_i}\,\ket{\Psi_i} \ket{\Phi_i} \, , \]
where \smash{``$\sum_{i=1}^k$''} serves to denote a quantum superposition on $k$ pure states.
If~the right-hand subsystem of \ket{\Psi} were measured by asking it to ``choose'' between one of the \ket{\Phi_i}'s,
each \ket{\Phi_i} would be chosen with probability \mbox{$\norm{\sqrt{p_i}}^2=p_i$}, leaving the unmeasured
left-hand subsystem in state~\ket{\Psi_i}.

Now, imagine that it were our friend the Daemon who prepared pure state \ket{\Psi} and measured its
right-hand subsystem. By~learning which \ket{\Phi_i} is obtained, with probability~$p_i$,
the Daemon would know in which \emph{pure} state \ket{\Psi_i} the unmeasured subsystem is.
If~the Daemon spits out this subsystem to the user, without revealing the result of the measurement,
the user receives a \emph{mixed} state corresponding to mixture~$\mathcal{E}$.
As~in Sect.~\ref{states}, this system is in a pure state for the Daemon and in a mixed state
for the user. The~beauty of this concept is that it~works even if the Daemon has not, in~fact,
measured the right-hand subsystem of the pure state it had created.
Even better, it still works if the Daemon has destroyed that right-hand subsystem,
inasmuch as a quantum state can be destroyed, to prevent any temptation to measure it later and
sell the answer to the user!
In~this case, the surviving quantum system would be in mixed state $\mathcal{E}$
not only for the user, but also for the Daemon.

The fact that any mixed state can be considered as the trace-out of one of its purifications
is the fundamental tenet of the Church of the Larger Hilbert Space, a term coined by
John Smolin because the formalism of quantum mechanics has pure quantum states
inhabit so-called Hilbert spaces and any mixed state can be thought of
as a subsystem from a pure state than lives in a \emph{larger} Hilbert space.

Everything that we have explained so far in this essay corresponds to strictly orthodox quantum mechanics
and no (serious) physicist would disagree with a single word from~it.
From this point on, however, we articulate our personal beliefs concerning the world in which we live,
which are admittedly very similar to the ``relative state''  formulation of quantum mechanics
put forward by \citet*{everett57} more than fifty years ago; see also \citet*{Byrne}.

The \emph{weak} Faithfuls in the Church of the Larger Hilbert Space believe in the fact
that any mixed state can be \emph{thought of} as the trace-out of some imaginary purification,
but this is only for mathematical convenience. In~fact, it is not possible to believe in the
predictions and formalism of quantum mechanics without being (at~least) a weak faithful since the
(mathematical) existence of a purification for any mixed state is a \emph{theorem}
that can be derived from first principles.

The \emph{strong} Faithfuls---among whom we stand---believe that to any mixed state
that actually exists, there corresponds somewhere in the Universe an appro\-priate purification.
This is an extremely far-reaching belief since it \mbox{implies} (among other things) that
the ``collapse of the wavefunction'', which orthodox quantum mechanics associates with
measurements, is but an illusion. In~fact, strong belief in the Church implies that quantum
mechan\-ics is strictly unitary and therefore reversible. If~we forget for simplicity the
necessity to apply relativistic corrections,
the Universe is ruled by one law only, known as Schr\"odinger's equation.
This equation is deterministic---even linear---and therefore so is the entire evolution of the Universe.

Let us consider for instance the simplest case of orthodox collapse of the wavefunction:
the measurement of a single diagonally-polarized photon (a~particle of light)
by an apparatus that distinguishes
between horizontal and vertical polarizations. For~definiteness, consider a calcite
crystal that splits an incoming light beam between horizontally and vertically polarized sub-beams
followed by two single-photon detectors (which we assume perfect for sake of the argument).
Any horizontally-polarized photon would cause one of the two detectors to react, whereas a
vertically-polarized photon would cause the other detector to react.
According to orthodox quantum mechanics, a diagonally-polarized photon would hit the crystal
and then continue in quantum superposition of both paths until it hits both detectors.
At~this point, one (and only one) of the detectors would ``see'' the photon and produce
a macroscopic effect that would be detectable by the (human) observer.
For~some, the phenomenon would become irreversible as soon as it has had a macroscopic
effect inside the detector; for others only when some observer becomes conscious of the outcome.

According to the Strong Church of the Larger Hilbert Space, neither is the case:
the diagonally-polarized photon
is in fact in an equal superposition of being horizontally and vertically polarized
(so~far, this is in strict accord\-ance with orthodox quantum mechanics) and
the crystal merely puts the photon in a superposition of both the horizontally and vertically polar\-ized
paths (still in accordance with orthodox quantum mechanics).
But when the photon hits both detectors, it becomes \emph{entangled} with them.
The composite system photon-detectors is now in an equal superposition
of the photon being horizontally-polarized and the horizontal-polarization detector having
\mbox{reacted} with the photon
being vertically-polarized and the vertical-polarization detector having reacted.
And when the observer looks at the detectors, he or she becomes entangled with
the photon-detector system so that now the photon-detecter-observer system
is in an equal superposition of the photon being horizontally-polarized,
the horizontal-polarization detector having reacted and the observer having
seen the horizontal-polarization detec\-tor reacting with
the same events corresponding to a vertically-polarized photon.

From this perspective, there is no collapse. The horizontal detection is as real as the vertical~one.
But any (human) observer becomes aware of only one outcome, and here lies the \emph{apparent}
paradox. If~indeed both events occur (in~quantum superposition), how come our experience
makes us (humans) believe that only one outcome (apparently chosen at random by Nature)
has actually occurred?
In~his groundbreaking paper, \citet*{everett57} proposed the following analogy:
\begin{quote}\sf
Arguments that the world picture presented by this theory is
contradicted by experience, because we are unaware of any branching
process, are like the criticism of the Copernican theory that the mobility
of the earth as a real physical fact is incompatible with the common sense
interpretation of nature because we feel no such motion.
In~both cases the argument fails when it is shown that the theory itself
predicts that our experience will be what it in fact is.
(In~the Copernican case the addition of Newtonian physics was required to be
able to show that the earth's inhabitants would be unaware of any motion of the earth.)
\end{quote}
In~other words, it is not because we (humans) cannot feel 
the Earth moving under our feet that it stands still at the centre of the Universe!
Similarly, it is not because we cannot feel
the Universal superposition that it does not exist.
According to the Strong Church of the Larger Hilbert Space,
the Earth as we feel it has but a tiny amplitude in the Universal wavefunction,
and each one of us has an even tinier amplitude.
This perspective is very humbling indeed, much more so than accepting
the insignificance of the Earth within the classical Universe,
but this is nevertheless the perspective in which we most passionately believe.

It remains to see how the Church of the Larger Hilbert Space can \mbox{account} for the
phenomenon described in Sect.~\ref{entanglement} when we discussed the
measurement of far-apart entangled particles\,%
\footnote{Of course, we must account for all the nonclassical correlations that violate
various forms of Bell inequalities, not only for the (classically explicable) fact that two particles in the singlet state
will always give opposite answers when subjected to the same measurement.
This paragraph can be adapted \emph{mutatis mutandis} to any pair of measurements,
including POVMs, on an arbitrary bipartite entangled state, as well as to similar scenarios
for multipartite entanglement.}.
Consider again two particles in the singlet state (Eq.~\ref{EPR}) and assume that they are both subjected
to the same measurement, which asks them to ``choose'' between orthogonal states
\ket{\psi} and~\ket{\phi}. We~can think of the two particles as being in the state
given by Eq.~(\ref{EPR2}), which once again is mathematically equivalent to Eq.~(\ref{EPR}), and initially the measurement apparatuses have not reacted,
so that they are not entangled with the particles.
The joint state of the apparatus-particle-particle-apparatus system can therefore be described as
\begin{equation}
\textstyle \ket{\mbox{?}} \left( \oosrtwod \ket{\psi}\ket{\phi} -
               \oosrtwod \ket{\phi}\ket{\psi} \right) \ket{\mbox{?}} \, ,
\end{equation}
where \ket{\mbox{?}} represents a measurement apparatus that has not yet reacted.
This is mathematically equivalent to
\begin{equation}
\textstyle \oosrtwod \ket{\mbox{?}}\ket{\psi}\ket{\phi}\ket{\mbox{?}} -
               \oosrtwod \ket{\mbox{?}}\ket{\phi}\ket{\psi}\ket{\mbox{?}} \, .
\end{equation}

Let us say without loss of generality that the particle on the left is measured first.
According to the Church of the Larger Hilbert Space,
this has the effect
of entangling it with its measurement apparatus. However (and contrary to the teachings of standard quantum
mechanics), the two particles remain entangled, albeit no longer in the singlet state.
Now, the joint state of the complete system has (unitarily) evolved to
\begin{equation}
\textstyle \left( \oosrtwod \ket{\Psi}\ket{\psi}\ket{\phi} -
               \oosrtwod \ket{\Phi}\ket{\phi}\ket{\psi} \right) \ket{\mbox{?}} \, ,
\end{equation}
where \ket{\Psi} (resp.~\ket{\Phi}) represents the state of a measurement apparatus that has registered
a particle in state \ket{\psi} (resp.~\ket{\phi}). Note that the apparatus on the right is still unentangled with
the other systems under consideration.
Finally, when the particle on the right is measured, the system evolves to
\begin{equation}\label{measured}
\textstyle \oosrtwod \ket{\Psi}\ket{\psi}\ket{\phi}\ket{\Phi} -
               \oosrtwod \ket{\Phi}\ket{\phi}\ket{\psi}\ket{\Psi} \, .
\end{equation}
At this point, if we trace out the two particles, the detectors are left in \emph{mixed} state
\begin{equation}
\textstyle  \{ ( \ket{\Psi}\ket{\Phi},\pbfrac12), (\ket{\Phi}\ket{\Psi},\pbfrac12) \}  ,
\end{equation}
which is exactly as it should: they have produced random but complementary outcomes.
Naturally, we could also involve two human observers in this scenario.
If~we had, they would enter the macroscopic entangled state of Eq.~(\ref{measured});
at~that point, they would in a superposition of having seen the two possible complementary
sets of outcomes, but they would be blissfully unaware of~this.

As~an amusing anecdote, we cannot resist mentioning the (real-life!)\ venture called
\textsf{cheap universes}.\,%
\footnote{\url{http://www.cheapuniverses.com}, accessed on 29 February 2012.}
For~a mere \$$3.95$, or unlimited use for \$$1.99$ on an iPhone, you can select two courses of action (such as ``I~shall either go on a
hike, or I~shall take a nice hot bath'') and ask \textsf{cheap universes} to make a purely quantum
choice between the two alternatives.\,%
\footnote{Specifically, \textsf{cheap universes} uses a commercial device called \textsf{QUANTIS},
available from \textsf{ID~Quantique}, in which ``photons are sent one by one onto a semi-transparent
mirror and detected; the exclusive events (reflection/transmission) are associated to `0'/`1' bit values''.
See~\mbox{\url{http://www.idquantique.com/true-random-number-generator/products-overview.html  
}}, accessed on 29~February~20102. According to our example, we would associate outcome 0 with
``I~shall go on a hike'' and outcome 1 with ``I~shall take a nice hot bath''.}
Provided you have self-pledged to obey the outcome,
you may proceed lightheartedly because you know that you are also performing the other action
in the Universal wavefunction.
Indeed, should the consequences of having indulged in a nice hot bath turn out to be disastrous,
you can take comfort in knowing that you have \emph{also} gone on a hike and hope that
this was indeed the path to happiness.
Sounds crazy? Not to~us!

The Strong Church of the Larger Hilbert Space is different from (but~not incompatible with)
the so-called Many-World Interpretation of Quantum Mechan\-ics (usually associated with the
name of Everett) in the sense that we believe in a single Universe---not~in the ``Multiverse''
advocated by the Many-World Inter\-pre\-ta\-tion followers---but~one
in which quantum mechanics rules at face value:
We~(poor humans) perceive only what we call the classical states,
but arbitrarily complex superpositions of them do in fact exist in reality.

\section{Can locality be restored outside of the Church?}\label{ontic}

Einstein thought that quantum mechanics must be \emph{incomplete} because
it did not fulfil his wish for a \emph{local} and \emph{realistic} theory.
We~distinguish between \emph{strong} realism, according to which
any property of a physical system registered by a measurement apparatus
(or~by any other process by which the system is observed) existed
prior to the measurement, so~that the apparatus merely reveals what was already there,
and \emph{weak} \mbox{realism}, according to which a physical system can respond
probabilistically to a measurement apparatus, but the probability distribution of
the possible outcomes exists prior to the measurement.
For instance, the diagonal polarization of a horizontally-polarized photon exhibits
weak realism according to quantum mechanics \mbox{because} its measurement behaves
randomly, yet with well-defined probabilities (in~this case with equal probability
of registering a $+45^\circ$ or a $-45^\circ$ outcome). 

Similarly, we distinguish between \emph{strong} locality, according to which
no action performed at point~\A{} can have an effect on point~\B{} faster than
the time it takes light to go from \A{} to~\B,
and \emph{weak} locality, according to which there can be no \emph{observable}
such effect.
As~we have seen already, if two particles are jointly in the singlet state (Eq.~\ref{EPR})
and if one is measured, yielding outcome \ket0, the other particle behaves as if its
state had instantaneously changed from being half a singlet to pure state~\ket1,
no matter how far apart the two particles are.
Even though this phenomenon \emph{seems} to violate strong locality
(we~shall come back on this issue below and in the next section),
it is important to understand that it does
\emph{not} violate weak locality because the instantaneous effect (if~it exists)
cannot be detected by any process allowed by quantum mechanics.
Taking account of the special theory of relativity, violations of weak locality would enable reversals in causality
(effects could precede causes), whereas violations of strong locality have no such spectacular consequences.
Fortunately, quantum mechanics does not allow \emph{any} violation of weak locality.
From now on, ``locality'' will be understood to mean ``strong locality'' unless specified otherwise
 
For a strong faithful in the Church of the Larger Hilbert Space, the issue of locality can take different flavours.
At~one extreme, the wavefunction is the one and only reality and the question does not even make sense.
The universe \emph{is} in a massive superposition and anything that appears to involve a random quantum
choice in one branch of the superposition ``simply'' makes the universal superposition more complicated;
the issue of locality does not even spring~up. This position is often considered to be a ``cop~out'' by those
who are not faithfuls of the Church, who think that believers are simply avoiding the issue rather than
trying to explain~it. The other extreme among the faithfuls is populated by the advocates of the
many-world interpretation of quantum mechanics,
some of whom consider that the entire world splits up each time a random quantum choice appears
to be made. Such a split is highly nonlocal if it is instantaneous.
In~the next section, we shall present our \emph{parallel lives} interpretation of
the Church of the Larger Hilbert Space, which is fully compatible with locality.
In~the rest of this section, however, we shall step outside of the Church and attempt
to reconcile locality with quantum mechanics while denying the possibility
for macroscopic objects (such as human observers) to enter into a superposition.

The great discovery of \citet*{bell64}, or~so it seems, is that the predictions of quantum mechanics
are incompatible with any possible strongly local and weakly realistic theory of the world.\,%
\footnote{To be historically more accurate, \citeauthor*{bell64}'s original \citeyear{bell64} paper was concerned
with strong realism only, but it can be strengthened to take account of weak realism.}
Since quantum mechanics has been vindicated by increasingly sophisticated
experiments \citep*[etc.]{clauser,aspect1,aspect2,aspect3,spacelike,spooky},
most physicists infer that there is no other choice but to forego locality.
However, we beg to disagree on the inevitability of this conclusion.
If~the world cannot be simultaneously local and realistic, could locality be restored
at the expense of realism?

One may attempt to achieve this by accepting that
the state of a quantum system is fundamentally \emph{subjective}
(or~to be technically more exact, \emph{epistemic}).
For~instance, the same particle can be in one state for one observer
and in a different state for another.
\emph{And both observers can be perfectly correct about the state of the particle!}
However, they must have \emph{compatible} beliefs in the sense that there must
exist at least one pure state that is excluded by neither observer.\,%
\footnote{To be technically exact and much more general, there must exist at least one \emph{ontic} state
compatible with both \emph{epistemic} beliefs, unless we are ready to accept that there is no
underlying reality at all \citep*{PBR}.}

For~sake of the argument, consider again a quantum system in the singlet state (Eq.~\ref{EPR})
so that the two particles are arbitrarily far apart, say at points~\A{} and~\B, which are inhabited
by Alice and Bob, respectively. We~have seen that the state of either
particle can be described locally by mixture $\mathcal{E}_1$ from Eq.~(\ref{comp-mixed}).
To~stress that we are not talking about the \emph{specific} mixture of pure states
explicit in~$\mathcal{E}_1$ (since the state of these particles can just as well be described
by mixtures $\mathcal{E}_2$ or~$\mathcal{E}_3$), let us denote the corresponding
density matrix by~$\rho$, which is uniquely defined.

Consider what happens if Alice measures her particle in the computational basis and
obtains (say) outcome~\ket0.
Then, assuming Bob has not interacted with his particle, Alice \emph{knows} that Bob's particle
is no longer in mixed state~$\rho$: now it is in state~\ket1.
But for Bob, nothing has changed! His~particle was in state~$\rho $ before Alice's measurement
and it \emph{still} is in this same state immediately after the far-away measurement.
In~other words, the particle at point~\B{} is in state~\ket1 for Alice and in state~$\rho $ for Bob,
\emph{and both observers are correct} in their assertions concerning the \emph{same} particle.
This is reminiscent of the proverbial Indian story of the blind men and an elephant.\,%
\footnote{\url{http://en.wikipedia.org/wiki/Blind_men_and_an_elephant}, accessed on 29~February~2012.}

The effect of Alice's measurement \emph{can} propagate to Bob,
but \emph{only} if a \emph{classical} message transits between them.
However, such a message cannot travel faster than at the speed of light.
It~follows that there is no faster-than-light change in the state of the particle at point~\B{},
as~seen by Bob from that point.
More generally, no operation performed at any point in space can have
an instantaneous \emph{observable} effect on any other point.
Seen this way, no cause can have an effect faster than at the speed of light,
causality is not violated,
and Einstein can rest in peace.

Naturally, it \emph{is} possible for an observer to be wrong about the state of a particle.
For instance, if Alice prepares a particle in state \ket0 and sends it to Bob, who is far away,
and if Bob subjects it to a Hadamard transformation without telling Alice, then Alice
may think that the particle is still in state \ket0 and be wrong since it is now in state~$\H\ket0$.
However, this is not in contradiction with the above: 
It~is not because the same particle can be in two different states according to two
different observers and that both can be correct that
anybody who has some opinion about the state of a quantum system is
necessarily right!
For Alice to know the state of a far-away particle, even subjectively, she must
know what has happened to it after it left her hands.
We~shall therefore consider for simplicity a bipartite scenario in which each party
knows what the other party is doing.

Can we completely restore locality at the expense of realism with this line of approach?
Unfortunately, there is a serious problem.
Consider again the case of Alice and Bob sharing a singlet state and of Alice measuring her particle.
We~argued above that Alice and Bob can both be correct if Alice thinks of Bob's particle as being in
state \ket1 whereas Bob thinks of it as being in mixed state~$\rho$.
But now, if Bob decides to measure his particle, and if indeed his belief that it is in state~$\rho$
were correct, there would be no \emph{local} reason that would prevent him from registering outcome~\ket0,
which is indeed possible when measuring~$\rho$ since it can be thought of
as mixed state $\mathcal{E}_1$ from Eq.~\ref{comp-mixed}.
This would no longer be compatible with Alice's belief that Bob's particle is in state \ket1,
even though each party knows what the other is doing.
Furthermore, if they meet in the future and compare notes, Alice and Bob will register correlations that are not
in accordance with quantum mechanics.
(Please remember that this section is written under the assumption that neither Alice nor Bob can
be in a superposition of having seen both results.)

Does it follow that quantum mechanics cannot be explained by a local theory even if
we are willing to forego realism?
It~turns out that we can have our cake and eat it too, provided we reintegrate the Church of the Larger Hilbert Space.
Contrary to the prevalent belief, the laws of nature can be simultaneously local and realistic,
and yet obey all the predictions of quantum mechanics.
In~order to reconcile this claim with Bell's impossibility proof,
please consider the quotation of Bell's at the opening of this essay and read on.

\section{Parallel Lives}\label{parallel}

We shall fully develop our \emph{parallel lives} theory for quantum mechanics in a subsequent paper.
Here, for simplicity, we explain how local realism can be consistent with bipartite correlations
that are usually considered to be even more nonlocal than those allowed by quantum mechanics.
Specifically, we consider the so-called \emph{nonlocal box} introduced by \citet*{PRbox}, which we
illustrate with a tale that takes place in an imaginary world, i.e.~in a~toy model of an alternative universe.
Our universe follows Einstein's special theory of relativity so that it is possible to assert, according
to the principle of weak locality, that some events cannot influence the outcome of other
far-away events that are sufficiently simultaneous.

Imagine two inhabitants of this universe, Alice and Bob, who travel very far apart in their spaceships.
Each one of them is carrying a box that features two buttons, labelled 0 and~1,
and two lights, one green and one red.
Once they are sufficiently distant,  Alice and Bob independently flip fair coins to decide
which button to push on their boxes, which causes one light to flash on each box.
The experiment is performed with sufficient simultaneity that Alice's box cannot know
the result of Bob's coin flip (hence the input to Bob's box) before it has to flash its own light, even if a signal travelling at
the speed of light left Bob's spaceship at the flip of his coin toss to inform Alice's box
of the outcome, and vice versa.\,%
\footnote{We are implicitly ruling out the local realistic theory of \emph{superdeterminism} here,
accord\-ing to which
there is no way to prevent the boxes from knowing which button is pressed on the other box,
not because a signal travels quickly enough between the boxes, but because everything
being deterministic, each box knows everything about the future, including which buttons will
be pushed anywhere in the universe.\\ See \url{http://en.wikipedia.org/wiki/Superdeterminism}, accessed on 29 February 2012.} 

After several instances of this experiment, Alice and Bob meet again to compare their results.
They discover to their amazement that they saw different colours when and only when they
had both pushed the ``1'' button.
In~a local classical world that denies the Church of the Larger Hilbert Space,
in which Alice and Bob cannot enter into a superposition (as~in the previous section), it~is easy to see that such boxes cannot exist.
More precisely, the best box that can be built cannot produce such results with a probability better 75\%.
In~a quantum-mechanical world, we can do better by the magic of entanglement, but the success probability cannot exceed
\mbox{$\cos^2\pi/8 \approx 85\%$} \citep*{tsirelson}, hence our imaginary world is not ruled
by quantum mechan\-ics either.
This is fine: remember that the purpose of this scenario is not to suggest a model of our world,
but rather to show that it is possible in a local realistic world to violate a Bell inequality.

What is the trick? Imagine that each spaceship lives inside a bubble.
When Alice pushes one button on her box, her bubble splits into two parallel bubbles.
Each bubble contains a copy of the spaceship and its inhabitant.
Inside one bubble, Alice has seen the red light flash on her box;
inside the other bubble, she has seen the green light flash.
From now on, the two bubbles are living parallel lives.
They cannot interact between themselves in any way and will never meet again.
The same phenomenon takes place when Bob pushes one button on his box.
Please note that Alice's action has strictly no instantaneous influence on Bob's bubble
(or bubbles if he has already manipulated his box): this splitting into parallel lives
is a strictly local phenomenon.

Let us consider what happens if Alice and Bob, each of whom now lives inside two parallel bubbles
although they cannot feel~it in any way\,%
\footnote{Remember \citeauthor*{everett57}'s analogy with medieval criticism of the Copernican theory
concerning the fact that we cannot feel the Earth move under our feet.}
decide to travel towards each other and meet again.
(A~similar scenario can be involved if they decide to use classical communication in order to compare notes,
rather than travelling.)
This is where magic\,\footnote{Remember Arthur C. Clarke's Third Law:
``Any sufficiently advanced technology is indistinguishable from magic''!}
takes place: Each of the two bubbles that contains
Alice is allowed to interact and see only a single bubble that
contains Bob, namely the bubble that satisfies the conditions
described above.
Note that such a perfect matching is always possible.
Furthermore, each bubble can ``know'' with which other bubble to interact
provided it keeps a (local) memory of which button was pressed and which light flashed.
In~this way, each copy of Alice and Bob
will be under the illusion of correlations that ``emerge from outside space-time'' \citep*{nicolas}. 
Yet~these correlations take place fully within space and time, in a completely local realistic universe.

Let us stress again that we are not claiming that our universe actually works as described above,
because it does not, according to \citet*{tsirelson}.
Our~point is that it is generally recognized that nonlocal boxes of the sort we have described
cannot exist in any local realistic world, \emph{and this is false} according to our toy model.
To~be more dramatic, consider Bell's theorem, or more precisely its best-known incarnation,
the CHSH inequality due to \citet*{chsh69}.
This inequality states that
``in~any classical theory [\ldots] a~particular combination of correlations\,%
\footnote{Specifically, $E(A,B)+E(A,B')+E(A',B)-E(A',B')$; for detail, please
see Eqs.~(1) and~(2) from
\citet*{PRbox}.}
lies between $-2$ and~$2$''
\citep*{PRbox}.
The original purpose of this inequal\-ity is that it is violated by quantum mechanics since the
same ``particular combination of correlations'' is predicted to be equal to $2\sqrt2$,
hence quantum mechanics cannot be explained by a ``classical theory'' of the sort considered
by \citeauthor*{chsh69} to derive their inequality.
As~demonstrated by \citet*{PRbox}, this combination can as large as~$4$ without violating weak locality,
and indeed it is equal to~$4$ in our toy model of the world.

Have we uncovered a fundamental mistake in the paper of \citeauthor*{chsh69}?
Not at all!
Bell's inequalities (including CHSH and those from \citeauthor*{bell64}'s original \citeyear{bell64} paper) are proved, indeed correctly, under the assump\-tion
that the classical world is a theory of \emph{local hidden variables}.
The confusion comes from the fact that this has been widely misinterpreted to mean that quantum mechanics
rules out any local realistic explanation of the world.
For~instance, \citet*{NielsenChuang} wrote in their book:
``These two assumptions together are known as the assumptions of local \mbox{realism}. [\ldots]
The Bell ine\-qual\-ities show that at least one of these assumptions is not correct. [\ldots]
Bell's inequalities together with substantial experimental evidence now points to the
conclusion that either or both of locality and realism must be dropped from our view of the world.''
Note that \citeauthor*{NielsenChuang} consider here both locality and realism to be of the strong type,
but our parallel lives mechanism is purely deterministic, hence it is strongly realistic as well.

The virtue of our toy model is to demonstrate in an exceedingly simple way that
local realistic worlds can produce correlations that are demonstrably impossible
in any classical theory based on local hidden variables.
Therefore, it illustrates the importance of understanding the true meaning of Bell's theorem.
Nevertheless, it begs the question: what about quantum mechanics?
Can \emph{it} be explained in a local realistic parallel lives scenario?

It turns out that the idea of quantum mechanics being local and realistic in a theory analogous to parallel lives
was discovered in the twentieth century: it~can be traced back at least to \citet*{DeutschHayden}.
Similar ideas were intro\-duced subsequently by \citet*{Rubin} and \citet*{Blaylock}.
The article of \citeauthor*{DeutschHayden} focused on locality without precisely formulating definitions of
realism or what we have called parallel lives, but their mathematical structure was quite similar to what
we propose here.
Of~course, a complete reformulation of quantum mechan\-ics along these lines is significantly
more technical and complicated than what is needed to ``explain'' nonlocal boxes,
but the conclusion is that the Church of the Larger Hilbert Space can be interpreted to
provide a fully deter\-min\-istic, strongly \mbox{local} and strongly \mbox{realistic} interpretation of quantum mechanics,
Bell's theorem notwithstanding.
Indeed, 
this interpretation is not about parallel branches or parallel universes in a multiverse,
but rather
about parallel lives, which is a purely local phenomenon.

We are currently working on a followup article that will provide much
more detail about our parallel lives theory.

\section{Free Will?}\label{freewill}

At this point, the fundamental question is ``Can a purely deterministic quantum theory give rise to at least the
\emph{illusion} of nondeterminism, randomness, probabilities, and ultimately can free will emerge from
such a theory''? Please note that this section is written at the first person as it reflects solely the opinion
of the first author.  The second author resolutely does not believe in free will and therefore
his position is that neither determinism nor randomness would be able to enable~it.

I~cannot answer in a definitive way the question asked at the beginning of this section.
Certainly, I~acknowledge the difficulty of
deriving the emergence of probabilities as mathematically inevitable from a quantum Universe
in which all events occur unitarily according to the Church of the Larger Hilbert Space \citep*{kent}.
However, we are faced with \emph{exactly} the same difficulty if the collapse of the wave function does occur,
or even in a purely classical world \citep*{PRR}.
I~also acknowledge the difficulty of deriving free will from probabilities, randomness and nondeterminism.
Nevertheless, I~am inhabited by an unshakable
belief that free will, \emph{if} it exists, cannot have another origin, with apologies to the compatibilists.

In his own essay for our meeting,
\citet*{nicolas} expresses his view that the Many-World Interpretation of
Quantum Mechanics ``leaves no space for free will''.
I~suspect that he would have the same opinion concerning
the Strong Church of the Larger Hilbert Space.
[He~also maintains that free will is not incompatible with the deterministic physics of Newton,
but I~fail to understand how classical physics could escape the ``\emph{intelligence}'' of \citet*{Laplace}, for which
``nothing would be uncertain and the future, as the past, would be present to its eyes''.]
In~any case, I~admit that
\citeauthor*{nicolas} may be right, but my most fundamental disagreement lies deeper than physics or mathematics when he says:
``I~enjoy free will much more than I~know anything about physics''.
I~respect his opinion but my personal position is that I~would prefer to live in a world
without free will rather than in one in which the wavefunction collapses nonunitarily.
After all, lack of free will in a deterministic universe does not deprive us
from our capacity to experience surprise and find wonder in the world,
because we cannot calculate, and hence predict, the future.
But of course, whether or not free will exists, it does not extend to the point of
letting each one of us choose in which of these two universes we actually live!

Perhaps \textsf{cheap universes} is our ultimate window on free will.
Provided we firmly decide to follow whichever course of action it chooses for us,
we are free to populate both branches of the Universal superposition.
In~whichever branch we perceive ourselves to be, we have made the free choice of letting
quantum phenomena decide for~us.
Of~course, I~am not seriously suggesting that free will did not exist until the inception
of \textsf{cheap universes}, just as \citet*{bell90} was not serious when he asked
if ``the wavefunction of the world [was]
waiting to jump for thousands of millions of years until a
single-celled living creature appeared? Or~did it have to
wait a little longer, for some better qualified system\ldots{}
with a PhD?''!

\section{Conclusion}\label{conc}

In this essay, we have penned down for the first time our beliefs concerning
the Universe in which we live, even though one of us (Brassard) has been inhabited by these thoughts for several decades.
The more time goes by, the more convinced we are that they constitute the most rational
explanation for our quantum world.
We~reject violently the notion that there would be a quantum-classical boundary
and that physics is discontinuous, with a revers\-ible (even unitary) evolution at the microscopic level
but an irreversible collapse at the macroscopic level of measurements.
It~may be that free will can at best be an illusion in a world ruled by the Strong Church
of the Larger Hilbert Space because every time you think that you make a
decision (provided you use the services of \textsf{cheap universes} or some other source of
true quantum randomness to make your choices),
you also make the complementary decision in the Universal superposition.
However, what does it matter if free will does not truly exist, provided the illusion
is perfect?\,\footnote{Seriously, we would not want to live in the Matrix imagined by the Wachowski brothers,
no matter how perfect is the simulation. So,~perhaps we \emph{do} care after all!}

We~give the last words of wisdom to \citet*{bell90}, who ended his Summary of ``Against `measurement'\,'' by:
\begin{quote}\sf
I mean [\ldots] by ÔseriousÕ, that ÔapparatusÕ should not be separated off from the rest of the world into black boxes,
as if it were not made of atoms and not ruled by quantum mechanics.
\end{quote}
Perhaps it's all nonsense, \emph{E~pur si muove!}

\section*{Acknowledgements}

G.\,B.\ expresses his unbounded gratitude to Charles H.~Bennett, who introduced him to the wonders of quantum mechanics
more than thirty years~ago. In~a very real sense, he can trace back most of his beliefs to Bennett's \mbox{patient} preaching.
He~is also deeply grateful to Christopher Fuchs, his malt brother, with whom he has had most fascinating
discussions concerning the foundations of quantum mechanics.
The above acknowledgements should not be construed into saying that Bennett and Fuchs would
agree with all that has been written here!

We also benefitted from several crucial discussions with too many wonderful people to list them here
even though they all helped shape our beliefs, but we must mention at least
Jeffrey Bub,
Claude Cr\'epeau,
Patrick Hayden,
Nicolas Gisin,
Adrian Kent,
Nathaniel David Mermin,
the late Asher Peres
and Alain Tapp.
Even though G.\,B.\ has lived with these ideas for several decades, and many people throughout
the years have asked him if he had anything written about~it, who knows how much longer it would have been
before he wrote them down, if ever, without the opportunity and motivation provided by Antoine Suarez and his
2010 meeting in Barcelona on \textit{Is~Science Compatible with Our Desire for Freedom?}\,%
\footnote{\url{http://www.socialtrendsinstitute.org/Activities/Bioethics/Is-Science-}\\\url{Compatible-with-Our-Desire-for-Freedom.axd}, accessed 29 February 2012.}

G.\,B.\ is supported in part by
the Natural Sciences and Engineering \mbox{Research} Council of Canada,
the Canada Research Chair program,
the Canadian Insti\-tute for Advanced Research
and the Institut transdisciplinaire d'informatique quantique.

\end{document}